\documentclass[prb,amsmath,amssymb,twocolumn,showpacs,floatfix,superscriptaddress]{revtex4-1}
\usepackage{graphicx}
\usepackage[english]{babel}
\usepackage{times}
\usepackage{dcolumn}
\usepackage[usenames,dvipsnames]{color}
\usepackage{units}
\newcommand{\beq}{\begin{equation}}
\newcommand{\eeq}{\end{equation}}
\newcommand{\bea}{\begin{eqnarray}}
\newcommand{\eea}{\end{eqnarray}}

\newcommand{\eps}{\epsilon}

\newcommand{\s}{\sigma}

\newcommand{\bfs}{\boldsymbol}

\begin{document}

\title{Quantum impurity coupled to Majorana edge fermions}

\author{Rok \v{Z}itko}

\affiliation{Jo\v{z}ef Stefan Institute, Jamova 39, SI-1000 Ljubljana,
Slovenia,\\
Faculty  of Mathematics and Physics, University of Ljubljana,
Jadranska 19, SI-1000 Ljubljana, Slovenia}

\author{Pascal Simon}

\affiliation{Laboratoire de Physique des Solides, CNRS UMR-8502, Universit\a'e Paris Sud, 91405 Orsay Cedex, France}

\date{\today}

\begin{abstract}
We study a quantum impurity coupled to the edge states of a two-dimensional
helical topological superconductor, i.e., to a pair of
counter propagating Majorana fermion edge channels with opposite spin
polarizations. For an impurity described by the Anderson impurity
model, we show that the problem maps onto a variant of the interacting
resonant two-level model which, in turn, maps onto the ferromagnetic
Kondo model. Both magnetic and non-magnetic impurities are considered.
For magnetic impurities, an analysis relying on bosonization and the numerical
renormalization group shows that the system flows to a fixed
point characterized by a residual $\ln 2$ entropy and anisotropic static and 
dynamical impurity magnetic susceptibilities. For
non-magnetic impurities, the system flows instead to a fixed point with no
residual entropy and we find diamagnetic impurity response at low temperatures.
We comment on the Schrieffer-Wolff transformation for problems with
non-standard conduction band continua and on the differences which arise when we 
describe the impurities by either Anderson or  Kondo impurity models.
\end{abstract}


\maketitle

\newcommand{\vc}[1]{{\mathbf{#1}}}
\newcommand{\vck}{\vc{k}}
\newcommand{\braket}[2]{\langle#1|#2\rangle}
\newcommand{\expv}[1]{\langle #1 \rangle}
\newcommand{\corr}[1]{\langle\langle #1 \rangle\rangle}
\newcommand{\bra}[1]{\langle #1 |}
\newcommand{\ket}[1]{| #1 \rangle}
\newcommand{\Tr}{\mathrm{Tr}}
\newcommand{\kor}[1]{\langle\langle #1 \rangle\rangle}
\newcommand{\degg}{^\circ}
\renewcommand{\Im}{\mathrm{Im}}
\renewcommand{\Re}{\mathrm{Re}}
\newcommand{\dtN}{{\dot N}}
\newcommand{\dtQ}{{\dot Q}}
\newcommand{\GG}{{\mathcal{G}}}
\newcommand{\atanh}{\mathrm{atanh}}
\newcommand{\sgn}{\mathrm{sgn}}

\section{Introduction}

Majorana fermionic operators are characterized by the relation
$\eta^\dag=\eta$, i.e., they represent particles which are their own
antiparticles. A pair of Majorana (real) operators can be combined
into one standard Dirac (complex) fermionic creation and one
corresponding annihilation operator which operate within the Fock
space of a single-level system. Majorana fermions can also be thought of as
an equal linear combination of particles and holes. In recent years, a number of
proposals have been advanced for physical realizations of
condensed-matter systems having low-energy excitation spectra which
can be formally described using Majorana fermion operators (see
Refs.~\onlinecite{wilczek2009},\onlinecite{stern2010},\onlinecite{franz2010},\onlinecite{qizhang},\onlinecite{hasan2010}
for recent reviews). In such interacting many-particle systems
Majorana fermions appear at low temperatures as emergent degrees of
freedom. In particular, it has been argued that topological
superconducting phases can be induced in topological insulators
\cite{kane2005f, fu2006f, bernevig2006f, moore2007f, bernevig2006sc,
konig2007d} by the proximity effect
\cite{fu2008majorana,fu2009inter,sau2010,sau2010prb,fu2009josephson,
tanaka2009nfis,qizhang,hasan2010}. These phases are characterized by
the presence of one or several Majorana excitation branches which are
localized along the perimeter of the two-dimensional superconducting
sheet. Such Majorana bands have energies inside the superconducting
energy gap. In the absence of defects they are, in fact, the only
excitations inside the gap. One distinguishes chiral topological
superconductors with a single Majorana continuum which propagates in a
unique direction (chirality) and helical topological superconductors
with two Majorana continua which counter propagate and have opposite
spin polarization (helicity or spin-momentum locking)
\cite{qizhang,qi2009,ryu2008,sato2009,qi2010tsc}.
While there is presently no confirmed physical realization of a helical
topological superconductor, several recent theoretical proposals are based on heterostructures made from topological insulators (such as Bi$_2$Se$_3$) and conventional superconductors (such as elemental superconducting metals).
By tuning magnetic doping of the topological insulating layer, it should be
in principle possible to obtain both chiral and helical topological superconductors \cite{qi2010tsc}.

An important issue in this field is the fate of the Majorana fermions
when electron-electron interactions are taken into account
\cite{gangadharaiah2011,stoudenmire} and the role of disorder
\cite{akhmerov11,potter11,brouwer11}, in particular that of impurities
\cite{shindou2010}. Dilute concentrations of impurities would
lead to various measurable low-temperature anomalies due to the Kondo effect, 
while higher concentrations could even destabilize the edge states and qualitatively affect the behavior of the system.
It is thus of considerable interest to accurately
analyze the behavior of interacting impurities embedded near the edges
of topological superconductors (TSC). This is particularly important
because the doping of the topological insulator layer by magnetic impurities
has been proposed as one of the methods to tune their material properties
and to control their ground states. It is therefore likely that some
magnetic dopant atoms will be invariably present at the edge area of the TSC, 
where the Majorana states are localized.

Since the impurities hybridize with the conducting states which have half the degrees of freedom of Dirac fermions (electrons), it is expected that there will be a number of particularities compared to the behavior of standard impurity
models, such as the Anderson and Kondo impurity models. Furthermore,
it is conceivable that one could attach quantum dots to such materials
so that they would hybridize with the edge states of TSCs; such hybrid
structures can also be described using the same class of impurity
models.

In this work we study the Anderson impurity model with modified
conduction-band and hybridization terms which only involve half the
degrees of freedom of the standard model. In Sec.~\ref{sec2}, we
describe the model and perform mappings to other known models.  In
Sec.~\ref{sec:bos}, we use bosonization techniques in order to gain
some qualitative understanding about the expected behavior of the
model. In Sec.~\ref{sec3}, this analysis is complemented by a
numerical analysis using a reliable non-perturbative technique, the
numerical renormalization group (NRG), which allows to address both
thermodynamic and dynamic properties of the system. By help of these
two complementary studies, we show that the model admits two
qualitatively different types of low-temperature fixed points, one
associated with magnetic impurities (those with large
electron-electron repulsion and occupancy near half-filling) which is
characterized by $\ln 2$ residual entropy and anisotropic magnetic
response, and another associated with non-magnetic impurities (those
with small electron-electron repulsion and/or occupancy away from
half-filling) which is characterized by zero residual entropy and
diamagnetic response. This is different from the standard Anderson
impurity model where the low-temperature fixed point always belongs to
the same family of Fermi-liquid fixed points and there are no
discontinuities as a function of the model parameters (interaction
strength, level energy, hybridization strength), only the
quasiparticle scattering phase shifts are smoothly changed. In
Sec.~\ref{sec4}, we perform the
Schrieffer-Wolff transformation in the case of Majorana bands. We comment on
the relations between the Anderson and Kondo impurity models in this
generalized setting and compare our results of a magnetic impurity coupled to Majorana
edge states with a recent
study\cite{shindou2010} which uses instead a Kondo Hamiltonian description. 
In particular, we find different results for the dynamical magnetic response and observe
that the phase transition predicted in Ref.~\onlinecite{shindou2010} does not
show up when the impurity is instead properly described by  a Anderson model. This is due to
the fact that the phase transition predicted in  Ref.~\onlinecite{shindou2010} 
occurs for parameter values of a Kondo model which cannot arise from
the Schrieffer-Wolff transformation of an Anderson model.

\section{Model and mappings}
\label{sec2}

\begin{figure}[htbp]
\centering
\includegraphics[width=8cm]{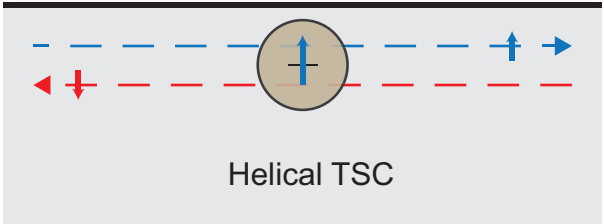}
\caption{(Color online) Schematic representation of the single
impurity problem studied in this work. A single impurity level
(circle) is hybridized with two continua of counter-propagating
Majorana fermions with opposite spin polarizations (dashed lines).}
\label{scheme}
\end{figure}

We study the single-impurity Anderson model (SIAM) consisting of an
impurity (single orbital) coupled by hybridization to the edge
states of a helical topological superconductor, i.e., to two continua
composed of counter propagating Majorana particles with opposite spin
polarization. The spin quantization axis is taken to be along the
$z$-axis; the rotation symmetry in the spin space is broken in this model.
The Hamiltonian can be obtained from the standard SIAM by projecting
out half of the conduction-band degrees of freedom. A schematic
representation is shown in Fig.~\ref{scheme}. The Hamiltonian is
\begin{equation}\label{eq0}
H=H_1+H_2+H_3,
\end{equation}
where $H_1$ is the impurity Hamiltonian
\begin{eqnarray}
H_1 &=& \delta \sum_\sigma (n_\sigma-1/2) + U (n_\uparrow-1/2)(n_\downarrow-1/2) \nonumber \\
&&+ B_x (1/2) (d^\dag_\downarrow d_\uparrow + d_\uparrow^\dag
d_\downarrow) \\
&&+ B_y (1/2) (id^\dag_\downarrow d_\uparrow -i d^\dag_\uparrow
d_\downarrow)\\
&&+ B_z (1/2) (n_\uparrow-n_\downarrow)\nonumber,
\end{eqnarray}
$H_2$ is the coupling Hamiltonian
\begin{eqnarray}
H_2 &=& {\sum_{k\sigma}} V_k 
\left(
\eta_{k\sigma} d_\sigma + d_\sigma^\dag \eta_{k\sigma} 
\right),
\end{eqnarray}
and $H_3$ is the conduction-band Hamiltonian for the Majorana modes
\begin{eqnarray}
H_3=\sum_{k>0} \left( v k \eta_{-k,\uparrow} \eta_{k,\uparrow}
+ v (-k) \eta_{-k,\downarrow} \eta_{k,\downarrow} \right).
\label{eq5}
\end{eqnarray}
The operators $d^\dag_\sigma$ with $\sigma=\uparrow,\downarrow$ create
an electron at the impurity level, and $n_\sigma=d^\dag_\sigma
d_\sigma$ is the electron number (occupancy) operator. The parameter
$\delta=\epsilon+U/2$ measures the departure from the particle-hole
symmetric point, $\epsilon$ is the on-site energy, $U$ is the
electron-electron repulsion parameter, while $B_x$, $B_y$, and $B_z$
are the components of the external magnetic field applied on the
impurity. The $\eta_{k,\sigma}$ are Majorana fermions operators satisfying
$\eta_{k,\sigma}^\dag=\eta_{-k,\sigma}$ and 
$\{\eta_{-k,\sigma},\eta_{k',\sigma'}\}=\delta_{kk'}\delta_{\sigma\sigma'}$.
$v$ is the Fermi velocity. The hybridization is given by the matrix
elements $V_k$. As common in the treatment of impurity models, we
neglect the $k$-dependence of $V_k$, i.e., $V_k \equiv V$; this is
permissible since we are mostly interested in the low-energy behavior
of the system. We furthermore assume that $V_k$ is real (see
Appendix~\ref{appa} for a generalization to complex $V_k$ where we show
that the phase of $V_k$ determines the orientation of the privileged axis,
to be discussed below, in the $xy$ plane that is perpendicular to the spin-quantization axis of Majorana bands, i.e., the $z$-axis). The
hybridization of the impurity with the conduction band is fully described by the
hybridization energy scale $\Gamma=\pi \rho |V|^2$, where $\rho$ is
the density of states in the conduction band. In this work, we
consider a constant density of states $\rho=1/\pi v$ so that $\Gamma$
is a constant independent of energy.

It is important to note that in writing the Hamiltonian in the form of
Eqs.~(\ref{eq0}-\ref{eq5}) we have omitted terms which describe the
coupling of the impurity to the bulk states of the superconductor
which have energies above the superconducting gap. These modes can
renormalize the effective parameters of the impurity Hamiltonian. For
simplicity, we assume that these high-energy modes have already been
integrated out and that the Hamiltonian $H$ is an effective low-energy
Hamiltonian valid on energy scales below the superconducting gap. The
bandwidth of the Majorana modes thus equals twice the superconducting
gap.

We now introduce the local Majorana operators $\eta_{i\sigma}$ through
\begin{equation}
\label{etaisig}
\begin{split}
d_{\sigma} = \frac{1}{\sqrt{2}} \left( \eta_{1\sigma} + i \eta_{2\sigma}
\right), 
&\quad
d^\dag_{\sigma} = \frac{1}{\sqrt{2}} \left( \eta_{1\sigma} - i
\eta_{2\sigma} \right),\\
\eta_{1\sigma} = \frac{1}{\sqrt{2}} \left(
d_\sigma+d_\sigma^\dag \right),
&\quad
\eta_{2\sigma} = \frac{1}{\sqrt{2}i} \left(
d_\sigma-d_\sigma^\dag \right),
\end{split}
\end{equation}
(note that we follow the normalization convention
$\eta_{i\sigma}^2=1/2$). The local spin and isospin operators can be
expressed as
\begin{equation}
\label{spinisospin}
\begin{split}
s_x &= \frac{1}{2} 
\left( d^\dag_\downarrow d_\uparrow + d^\dag_\uparrow d_\downarrow \right) =
\frac{i}{2} 
\left( \eta_{1\downarrow} \eta_{2\uparrow} + \eta_{1\uparrow}
\eta_{2\downarrow} \right) \\
s_y &= \frac{1}{2} 
\left( i d^\dag_\downarrow d_\uparrow - i d^\dag_\uparrow d_\downarrow \right) =
\frac{i}{2} 
\left( \eta_{1\downarrow} \eta_{1\uparrow} + \eta_{2\downarrow}
\eta_{2\uparrow} \right) \\
s_z &= \frac{1}{2} 
\left( d^\dag_\uparrow d_\uparrow - d^\dag_\downarrow d_\downarrow \right) =
\frac{i}{2} 
\left( -\eta_{1\downarrow} \eta_{2\downarrow} + \eta_{1\uparrow}
\eta_{2\uparrow} \right) \\
i_x &= \frac{1}{2} 
\left( d^\dag_\uparrow d^\dag_\downarrow + d_\downarrow d_\uparrow \right) =
\frac{i}{2} 
\left( \eta_{1\downarrow} \eta_{2\uparrow} - \eta_{1\uparrow}
\eta_{2\downarrow} \right) \\
i_y &= \frac{1}{2} 
\left( i d^\dag_\downarrow d^\dag_\uparrow + i d_\downarrow d_\uparrow \right) =
\frac{i}{2} 
\left( \eta_{1\downarrow} \eta_{1\uparrow} - \eta_{2\downarrow}
\eta_{2\uparrow} \right) \\
i_z &= \frac{1}{2} 
\left( d^\dag_\uparrow d_\uparrow + d^\dag_\downarrow d_\downarrow - 1 \right) =
\frac{i}{2} 
\left( \eta_{1\downarrow} \eta_{2\downarrow} + \eta_{1\uparrow}
\eta_{2\uparrow} \right) 
\end{split}
\end{equation}
We rewrite $H_1$ as
\begin{equation}
\begin{split}
H_1 &= \delta \sum_\sigma i \eta_{1\sigma} \eta_{2\sigma}
+ U \eta_{1\uparrow} \eta_{1\downarrow} \eta_{2\uparrow} \eta_{2\downarrow} \\
& + B_x(i/2)
\left( \eta_{1\downarrow} \eta_{2\uparrow} + \eta_{1\uparrow}
\eta_{2\downarrow} \right) \\
& +
B_y(i/2) \left( \eta_{1\downarrow} \eta_{1\uparrow} + 
\eta_{2\downarrow} \eta_{2\uparrow} \right) \\
& +
B_z(i/2) \left( \eta_{1\uparrow} \eta_{2\uparrow}-
\eta_{1\downarrow} \eta_{2\downarrow} \right),
\end{split}
\end{equation}
while
\begin{equation}
H_2 = \sum_{k\sigma} V
\sqrt{2}i \eta_{k\sigma} \eta_{2\sigma}.
\end{equation}
Only the two $\eta_{2\sigma}$ Majorana local modes are coupled to the
continuum (hybridized). For $\delta=B=U=0$, the two $\eta_{1\sigma}$
modes are fully decoupled from the rest of the system. For $U \neq 0$,
the $\eta_{1\sigma}$ modes interact via the quartic term with the
$\eta_{2\sigma}$ modes. As we show in the following, this brings about
non-trivial effects. For $\delta=B=0$ and finite $U$, we will see that
on the temperature scale $T_1=f(U,\Gamma)$ the fluctuations of the
$\eta_{2\sigma}$ modes are frozen out and that the entropy is reduced
by $2 \times (1/2)\ln2=\ln2$. The $\eta_{1\sigma}$ modes remain
active, and the system has $\ln 2$ residual entropy down to $T=0$. We
would therefore like to study the nature and the dynamics of this
residual degree of freedom. We find that
\begin{equation}
i\eta_{1\downarrow}\eta_{1\uparrow} = s_y+i_y.
\end{equation}
The residual degree of freedom thus corresponds to a mixed
spin-isospin mode associated with a linear combination of spin and
isospin projection operators along the $y$ direction (see
Appendix~\ref{appa} for a generalization to an arbitrary phase factor
in $V_k$). For this reason we expect
that the model has different magnetic response in the $y$ direction as
compared to the $x$ and $z$ directions. In the following, the
privileged $y$ direction will be referred to as the ``longitudinal''
direction, while the other two will be called ``transverse''
directions.

We now map the problem onto an interacting resonant two-level model.
For that purpose, we introduce new Dirac fermionic operators defined
by
\begin{equation}
\begin{split}
a_k &= \frac{1}{\sqrt{2}} \left( \eta_{k\uparrow} +
i \eta_{k\downarrow} \right), \\
a_k^\dag &= \frac{1}{\sqrt{2}} \left( \eta_{-k\uparrow} -
i \eta_{-k\downarrow} \right), \\
-i b &= \frac{1}{\sqrt{2}} \left( \eta_{2\uparrow}
+i \eta_{2\downarrow} \right), \\
f &= \frac{1}{\sqrt{2}} \left( \eta_{1\uparrow} + i \eta_{1\downarrow}
\right).
\end{split}
\end{equation}
In terms of these operators,
\begin{eqnarray}
\label{rl}
H_1 &=& \delta \left( b^\dag f + \text{H.c.} \right)
- U (n_b-1/2)(n_f-1/2) \nonumber\\
&&+B_x/2 \left( i b^\dag f^\dag + \text{H.c.} \right) 
+B_z/2 \left( b^\dag f^\dag + \text{H.c.} \right),\nonumber\\
&&+B_y/2 \left( 1-n_b-n_f \right) \\
H_2 &=& \sum_k V \sqrt{2} \left( a^\dag_k b + \text{H.c.} \right),\\
H_3&=&\sum_k v k \;a^\dag_{k}a_k.\label{rl1}
\end{eqnarray}
Notice that the summation in the last equation is indeed for all $k$.
Here $n_b=b^\dag b$ and $n_f=f^\dag f$ are the occupancy operators for
orbitals $b$ and $f$, respectively. This is a variant of the
interacting resonant two-level model for spinless Dirac fermions with
{\sl attractive} charge-charge coupling. The $b$ mode directly
hybridizes with the continuum, while the $f$ mode is fully decoupled
for $\delta=U=B=0$. We note the similarity between the $B_x$ and $B_z$
terms in Eq.~\eqref{rl} (they differ essentially only by a phase shift
of the operators), while $B_y$ has a qualitatively different form.
This is in line with the expected anisotropy of the system. The sum of
the terms in $B_x$ and $B_z$ can be compactly rewritten as $(B_\perp\,
b^\dag f^\dag+H.c.)/2$, where 
\begin{equation}
B_\perp=B_z+iB_x.
\end{equation}
The interesting behavior results for non-zero parameters $\delta$,
$U$, and/or $B$. For finite interaction $U$, we expect to observe
orthogonality catastrophe physics due to the effect of the occupancy
changes at the impurity level $f$ on the continuum. 

In order to capture the physics associated with this model, let us
first use a field-theoretical approach. This is the subject of the
next section.

\section{Bosonization approach}\label{sec:bos}

A Hamiltonian somewhat similar to Eqs.~(\ref{rl}-\ref{rl1}) has been
studied recently in the context of the charging of a narrow level in a
quantum dot capacitively coupled to a broader one \cite{goldstein2007,
karrasch2007, kashcheyevs2009, goldstein2010}. However, contrary to
the Hamiltonian in Eq. (\ref{rl}), the levels in the dot were
interacting via a repulsive Coulomb interaction, in which case the
system always has a non-degenerate ground state and is a conventional
Fermi liquid. Here, the interaction is attractive which, as we will
see, can significantly change the physics.

We start with the case $B_y=0$.  We follow the bosonization procedure
developed in Ref. \onlinecite{kashcheyevs2009} and repeat the main
steps for completeness. We first diagonalize the Hamiltonian $H_2+H_3$
which corresponds to a resonant-level model. The density of states of
the level $b$ corresponds to a Lorentzian of finite width $\Gamma$
centered around zero energy. If we assume $\delta,|B_\perp|\ll
\Gamma$, we can replace the density of states of the level $b$ by a
flat density \cite{kashcheyevs2009} for energies
$\omega\in[-\pi\Gamma/2,\pi\Gamma/2$] and $0$ elsewhere, with height
$1/(\pi\Gamma)$.  In this limit, the energy scale $\Gamma$ plays the
role of an effective bandwidth. Therefore, $H$ reads in the continuum
and low-energy limit
\begin{eqnarray}
\label{Hcont}
H&=&-iv\int\limits_{-\infty}^\infty \psi^\dag(x)\partial_x\psi(x)dx-Ua:\psi^\dag(0)\psi(0):(n_f-1/2)\nonumber\\
&&+\delta \sqrt{a}(\psi^\dag(0)f+H.c.)+B_\perp \sqrt{a}(\psi^\dag(0)f^\dag+H.c.), 
\end{eqnarray}
where $a\sim 2v/\Gamma$ is a new UV cut-off associated with the new effective bandwidth $\Gamma$, $\psi(x)$
is a right-moving field associated with the diagonalization of the low-energy limit of $H_2+H_3$.
The Hamiltonian in Eq. (\ref{Hcont}) can be treated using
bosonization.  We introduce a bosonic field
associated to the chiral fermionic one $\psi(x)\approx \frac{1}{2\pi
a}e^{-i\phi(x)}$, such that
\begin{eqnarray}
\label{Hbos}
H&=&v\int\limits_{-\infty}^\infty\frac{dx}{4\pi}\, (\nabla\phi)^2 
-v\frac{2\alpha_U}{\pi}\nabla\phi(0)(n_f-1/2)\\
&&+\frac{\delta}{\sqrt{2}}(e^{i\phi(0)}f+H.c.)+\frac{B_\perp }{\sqrt{2}}(e^{i\phi(0)}f^\dag+H.c.),\nonumber
\end{eqnarray}
where 
\begin{equation}
\alpha_U=\arctan(U/2\Gamma)
\end{equation}
is fixed so that the phase shift we obtain for $\delta=B_\perp=0$ in
the fermionic representation equals the one calculated in the bosonic representation
for a given occupation of the $f$ level.\cite{kashcheyevs2009} Next,
we apply the canonical transformation $H'=U^\dag H U$ with 
\begin{equation}\label{eq:U}
U=\exp[i(2\alpha_U/\pi)\phi(0)(n_f-1/2)],
\end{equation}
such that
\begin{eqnarray}
\label{Hbos1}
H'&=&v\int\limits_{-\infty}^\infty\frac{dx}{4\pi}\, (\nabla\phi)^2 +\frac{\delta}{\sqrt{2}}(e^{i\gamma_+\phi(0)}f+H.c.)\nonumber   \\
&&+\frac{B_\perp }{\sqrt{2}}(e^{i\gamma_{-}\phi(0)}f^\dag+H.c.),
\end{eqnarray}
where we have introduced \beq\label{eq:gamma} \gamma_\pm=1\pm
2\alpha_U/\pi. \eeq 
Let us first consider the following two situations: $\delta\ne
0,B_\perp=0$ and $\delta= 0,B_\perp\ne 0$. In both cases, the
Hamiltonian in Eq. (\ref{Hbos1}) maps to a bosonized formulation of
the anisotropic Kondo model
\begin{equation}\label{eq:HAK}
H_{AK}= \sum_{k\sigma} \epsilon_k c_{k\sigma}c_{k\sigma} 
+\sum_{kk'\sigma\sigma'} J_i {\bfs S}^i\bfs \sigma^i_{\s\s'}
c_{k\s}^\dag c_{k'\s'},
\end{equation}
where $\eps_k=vk$ and $\bfs \s^i$ are the Pauli matrices ($i=x,y,z$).
The  Kondo couplings $J_i$ are such that $J_x=J_y=J_\perp$.
The bosonization of (\ref{eq:HAK}), as derived in
Refs.~\onlinecite{schlottmann},\onlinecite{giamarchi1993}, provides the identification
\begin{equation}
\begin{split}\label{eq:ident1}
\delta &= J_\perp/\sqrt{8}\\
\gamma_{\pm} &= \sqrt{2}[1-\frac{2}{\pi}\arctan(\pi\rho J_z/4)],
\end{split}
\end{equation}
where $\rho$ is the density of states in the large-$\Gamma$ limit we
are considering. In the limit where $\Gamma$ is much larger than $U$,
we can linearize the $\arctan$ functions and estimate
\begin{equation}
\label{Jz}
J_z \approx \pi\Gamma(2-\sqrt{2})\mp\sqrt{2}U.
\end{equation}
The upper sign $(-)$ corresponds to the case $\delta\neq0, B_\perp=0$,
and the lower sign $(+)$ to the case $\delta=0, B_\perp\neq0$.
Let us now discuss both cases separately.

{\em Case $\delta\ne0$ and $B_\perp=0$}. -- We see from Eq. 
(\ref{Jz}) that a large $U\gg \Gamma$ enforces $J_z<0$, {\it
i.e.}, a ferromagnetic Kondo model, while a small $U$ leads to an
antiferromagnetic Kondo model. In the latter case, we can extract a
Kondo scale $T_K^+$ associated with the screening of the magnetic
moment. From the known results on the anisotropic Kondo
model,\cite{anderson70} we find \beq T_K^+\sim
\Gamma\left(\frac{\delta}{\Gamma}\right)^{\frac{2}{2-\gamma_+^2}},
\eeq with $\gamma_+$ defined in Eq. (\ref{eq:gamma}). We remind that
this result makes sense only for $J_z>0$ (small-$U$ case).

In order to estimate the low energy physics in the presence of both
$\delta$ and $U$ terms, we can rely on the lowest order renormalization group (RG) equations for the Kondo model\cite{giamarchi1993}
\bea
\frac{dJ_\perp}{dl}&=&\rho J_\perp J_z -\rho^2 J_\perp(J_\perp^2+J_z^2)/4\\
\frac{dJ_z}{dl}&=&\rho J_\perp^2 -\rho^2 J_zJ_\perp^2/2,
\eea
where $l$ is a running length scale.
In particular, if we neglect the cubic terms, this is the standard Kosterlitz-Thouless RG flow
where $J_z^2-J_\perp^2$ is a constant of the flow.
When $J_z<0$, the infrared fixed point depends on whether $|J_z(a)|$
is above or below the separatrix $J_\perp=-J_z$, assuming $J_\perp>0$.
If $|J_z(a)|>J_\perp(a)$, the fixed point is a strong-coupling one which corresponds to the antiferromagnetic Kondo fixed point and the formation of a Kondo singlet.
However, for $|J_z(a)|<J_\perp(a)$, the flow is driven toward a fixed line corresponding to $J_\perp\to 0$ and
$J^z\to J_z^*<0$ associated with the ferromagnetic Kondo model and therefore an unscreened free moment.
Although we have assumed $\Gamma$ to be  larger than other scales, the condition $J_\perp(a)>|J_z(a)|$
associated with a Fermi-liquid fixed point and a screened magnetic moment translates into 
\begin{equation}
\label{bozoni}
\delta > U/2-\frac{\pi(\sqrt{2}-1)}{2}\Gamma.
\end{equation}

{\em Case $\delta=0$ and $B_\perp\ne 0$}. -- This case is simpler since $J_z>0$ for all values of $U$.
Therefore, the low energy physics is described by a Fermi liquid fixed point with a screened magnetic moment
and we can define a Kondo scale $T_K^-$ by
\beq
\label{tkminus}
T_K^-\sim \Gamma\left(\frac{B_\perp}{\Gamma}\right)^{\frac{2}{2-\gamma_-^2}},
\eeq
with $\gamma_-$ defined in Eq. (\ref{eq:gamma}).

{\em General case}. -- When both $\delta$ and $B_\perp$ are different
from zero, the mapping to the anisotropic Kondo problem breaks down.
However, at small $U$ compared to $\Gamma$, both terms independently
favor a strong-coupling fixed point with a screening of the local
moment. Therefore, we expect the same behavior in the presence of both
terms. In the large-$U$ limit with $B_\perp=0$, a ferromagnetic Kondo
model is favored and a local moment is stabilized. The addition of a
small magnetic field $B_\perp$ around this fixed point is, however, a
relevant perturbation which destabilizes this behavior and a
non-magnetic fixed point is also expected.

In our analysis, we have considered $B_y=0$. However, we can extend
the previous bosonization analysis to models with $B_y\ne 0$. The term
$B_y$ has two effects: First, it provides an additional  potential
scattering term $-B_y a/2:\psi(0)\psi(0):$ in the continuum-limit
Hamiltonian in Eq. (\ref{Hcont}) which will modify the phase-shift
analysis. Second, it adds an extra term $-B_y (n_f-1/2)/2$ which
corresponds to a magnetic field of the form $-B_y S^z/2$ in the
equivalent Kondo Hamiltonian in Eq. (\ref{eq:HAK}). As discussed
above, the latter term would polarize a free local moment
and no residual entropy is expected at energy below this magnetic
field. 

\section{Numerical renormalization group analysis}
\label{sec3}

\begin{figure}[htbp]
\centering
\includegraphics[clip,width=8cm]{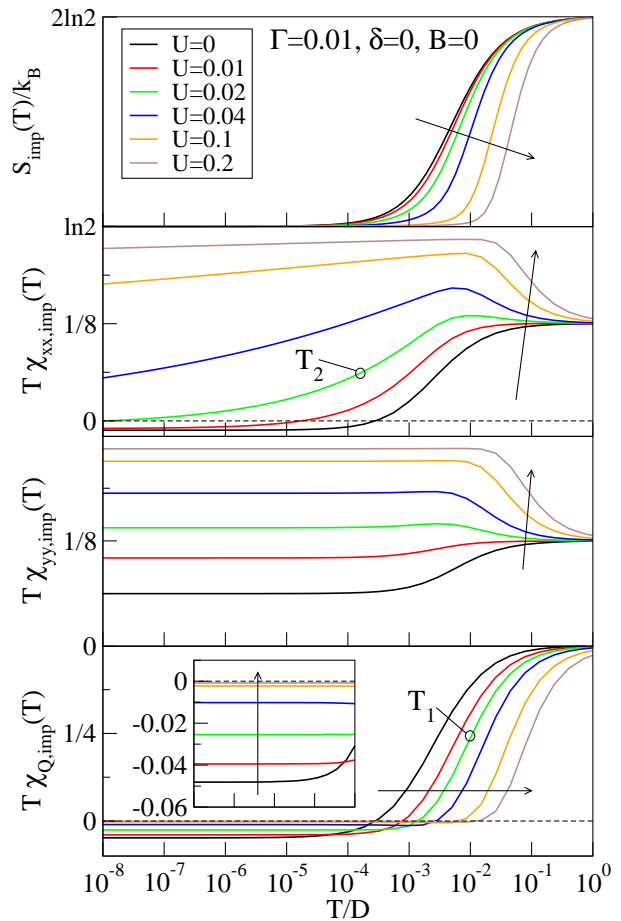}
\caption{(Color online) Thermodynamic properties (impurity entropy and
impurity susceptibilities) at the particle-hole symmetric point $\delta=0$ 
for a range of values of the
interaction strength $U$. The arrows indicate the direction of
increasing $U$. In the absence of the external magnetic field, one has
$\chi_{xx,\mathrm{imp}}=\chi_{zz,\mathrm{imp}}$ due to symmetry.
The out-of-diagonal magnetic susceptibilities are all zero.
On the curves for $U=0.02$, we label the positions of the
temperatures $T_1$ and $T_2$ (see the main text for their definitions).
}
\label{fig1}
\end{figure}

We now study the Hamiltonian in its original (non-transformed) form of
Eq.~\eqref{eq0} using the numerical renormalization group (NRG) method
\cite{wilson1975, krishna1980a, bulla2008}. This technique is
essentially an iterative exact diagonalization of an appropriately
discretized form of the Hamiltonian, where at each partial
diagonalization step the high-energy excitations are discarded
(``truncated'') and only the low-energy part of the spectrum is
retained in the subsequent calculation steps. This approximation is
appropriate because the matrix elements coupling levels with widely
different energy scales are small \cite{wilson1975}. The NRG method
has already been applied to problems involving an impurity coupled to
Majorana edge fermions; for details, see Ref.~\onlinecite{detection}.
(In this work, the NRG calculations have mostly been performed using
parameters $\Lambda=3$, $E_\mathrm{cutoff}=10$ or at most 5000 states,
whichever is lower, $N_z=4$, $\alpha=0.6$.)

\subsection{Thermodynamic properties: particle-hole symmetric case}

We first study the problem for a particular choice of $\delta=0$,
i.e., we focus on the particle-hole (p-h) symmetric problem. We will
show in the following that there is actually a finite region in the
$(\delta,U,\Gamma)$ parameter space where the problem flows to the
same type of the low-temperature fixed point as along the p-h
symmetric line, thus the results are more general. The impurity
entropy and thermal spin and charge susceptibilities are shown in
Fig.~\ref{fig1} for a range of the interaction strengths $U$,
including for the non-interacting $U=0$ limit. We remind the reader
that the impurity entropy quantifies the number of the effective degrees
of freedom $p$ on the impurity via 
\begin{equation}
S_\mathrm{imp}(T)=k_B \ln p(T).
\end{equation}
The impurity magnetic susceptibility corresponds to the effective
local magnetic moment as 
\begin{equation}
\mu_{i}^2(T)=T \chi_{ii,\mathrm{imp}}(T),
\end{equation}
where 
\begin{equation}
T\chi_{ij,\mathrm{imp}}(T)=\langle S_{i} S_{j} \rangle -
\langle S_{i} S_{j} \rangle_0,
\end{equation}
with $S_i$ the $i$-component of the total spin of the system. Here
$\langle \rangle_0$ denotes the expectation value for a system without
the impurity. Finally, the impurity charge susceptibility
$\chi_{Q,\mathrm{imp}}$ is defined through
\begin{equation}
T\chi_{Q,\mathrm{imp}}(T) = \langle Q^2 \rangle - \langle Q^2 \rangle_0
\end{equation}
where $Q$ is the total charge in the system. Since the
impurity susceptibility is defined as the difference between an expectation
value in the full system and the expectation value in the system
without the impurity (i.e., the impurity contribution to the total system susceptibility), the impurity susceptibilities can be negative. This is different from the local susceptibilities of the impurity, which are different quantities that are positive-definite.

In the high-temperature limit, the system behaves as the standard SIAM
in the same limit \cite{krishna1980a}: the impurity freely fluctuates
between all four possible states, thus the entropy is $\ln4$, while
the magnetic moment is $(0+1/4+1/4+0)/4=1/8$, since only two of the
four states have magnetic moment of $1/4$. The charge moment
$T\chi_{Q,\mathrm{imp}}$ is $1/2$ due to maximal charge fluctuations
on the impurity.

On the temperature scale $T_1 \approx \max(U,\Gamma)$, the directly
coupled Majorana modes $\eta_{2\sigma}$ are frozen out (or, which is
equivalent, the Dirac mode $b$ is frozen out) and we enter a $\ln2$
plateau in the impurity entropy which persists down to $T=0$.  The
freezing-out of the charge degrees of freedom is also reflected in the
effective charge moment $T \chi_{Q,\mathrm{imp}}(T)$, which on the
temperature scale $T_1$ drops to some small (negative) value and
remains constant down to $T=0$. The larger $U$ is, the larger is the
reduction of the charge fluctuations on the impurity, and the closer
does the charge moment approach to zero. We emphasize that in the
standard SIAM the effective charge susceptibility goes to zero at low
temperatures and it is always positive.

The residual $\ln 2$ entropy suggests that the other two Majorana
modes $\eta_{1\sigma}$ (i.e., the Dirac mode $f$) remain decoupled; this
is strictly true only for the non-interacting $U=0$ model, see
Eq.~\eqref{rl}, but an effective decoupling also occurs for $U \neq 0$. 
The magnetic susceptibility curves indicate, however, that the
behavior within the $\ln 2$ entropy plateau is not trivial, see the
two middle panels in Fig.~\ref{fig1}. The most striking feature is the
strongly anisotropic behavior: while the effective moments in the
transverse $x$ and $z$ directions are reduced (``screened'') at low
temperatures, the longitudinal ($y$-axis) effective moment reaches its
asymptotic value on the temperature scale of $T_1$ and remains
constant down to $T=0$. We also observe that the behavior is different
depending on the value of the ratio $U/\Gamma$. (We remind the
reader that this same ratio controls the stability of the non-magnetic
solution of the Hartree-Fock equations for the standard SIAM. For
$U/\pi\Gamma<1$ the paramagnetic solution is stable, while for
$U/\pi\Gamma>1$ the solution breaks the spin symmetry and the
Hartree-Fock approximation is no longer applicable. This ratio thus
determines whether the system is in the ``Kondo regime'' or not.)

\begin{figure}[htbp]
\centering
\includegraphics[width=7cm,clip]{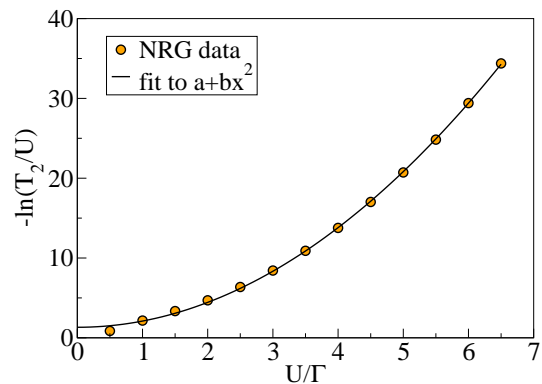}
\caption{(Color online) The temperature $T_2$ where the effective
transverse magnetic moment $\mu^2_x=T\chi_{xx,\mathrm{imp}}(T)$ is reduced to $1/8$.
The full curve is a fit which corresponds to Eq.~\eqref{tkfit}.}
\label{tk}
\end{figure}

Let us first consider the case of large $U$, i.e., $U/\Gamma
\gtrsim 1$, which corresponds to the magnetic regime. The transverse
magnetic susceptibility indicates a progressive reduction of the
effective impurity magnetic moment. This reduction occurs at some low
temperature scale $T_2$ which appears to be exponential in $U/\Gamma$,
similar to the exponential behavior of the Kondo temperature in the
standard SIAM. The scale $T_2$ may be defined, somewhat arbitrarily,
through $\mu^2_x(T_2)=\mu^2_x(T=\infty)/2=1/8$. The temperature $T_2$
is plotted as a function of $U/\Gamma$ in Fig.~\ref{tk}. We find an
excellent fit to the function
\begin{equation}
\label{tkfit}
T_2 = c_1 U \exp\left[-c_2 \left(\frac{U}{\Gamma}\right)^2 \right],
\end{equation}
where $c_1=3.73$ and $c_2=0.78 \approx \pi/4$. This is distinctly
different from the expression for the Kondo temperature in the
standard SIAM which takes the form $T_K \sim U \exp(-\pi U/8\Gamma)$.
The scale $T_2$ appears to be in many respects similar to the
low-energy scale studied in Ref.~\onlinecite{kashcheyevs2009},
although the dependence on the parameters is exponential rather than
algebraic. The physical mechanism which controls the scale $T_2$ is
not clear at present.

Note that at zero temperature the impurity spin is not screened in the
usual sense (which would be associated with a reduction of the
impurity entropy to zero due to the formation of a non-degenerate
spin-singlet Kondo state). In particular, the effective longitudinal
moment is still non-zero, thus the spin is maximally anisotropic; it
behaves as an Ising spin which cannot rotate away from the $y$
direction.

For small $U$ of order $\Gamma$, we find somewhat different behavior.
In this case, the scales $T_2$ and $T_1$ are not well separated: the
charge and spin susceptibilities are reduced simultaneously. Again,
the system has residual $\ln2$ entropy and there is residual
longitudinal magnetic moment.

In analogy with the standard SIAM, we find that the interaction
strength $U$ plays the role of the effective bandwidth
\cite{krishna1980a}, so that for constant $U/\Gamma$ ratio the
thermodynamic curves are simply shifted to lower energy scales
proportionally to $U$. Finally, it should also be remarked that at the
particle-hole symmetric point, the low-temperature stable fixed point
of the system is exactly the same irrespective of the values of the
parameters $U$ and $\Gamma$; again, a similar feature is also found in
the standard SIAM \cite{krishna1980a}.

\begin{figure}[htbp]
\centering
\includegraphics[clip,width=8cm]{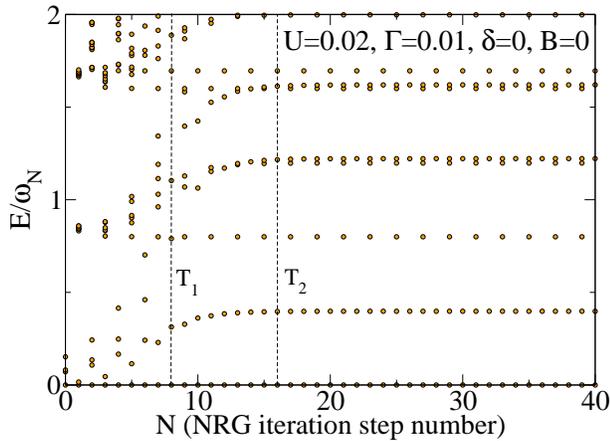}
\caption{Renormalization-group flow diagram. The NRG iteration step
numbers corresponding to the temperature scales $T_1$ and $T_2$ are
indicated.}
\label{spag}
\end{figure}

It is interesting to note that the second scale $T_2$ does not very clearly appear
in the renormalization-group energy-level flow diagram,
Fig.~\ref{spag}, which is consistent with the constancy of
$S_\mathrm{imp}$ at low temperatures. The fact that the susceptibility
curves nevertheless exhibit temperature dependence then implies that
it is the nature of the low-energy excitations which changes with the
temperature. There are no conserved quantum numbers in the problem,
thus the information about the spin susceptibility is contained in the
matrix elements, such as $\bra{i} S_z \ket{j}$, between the energy
eigenstates. These matrix elements do have an RG flow. At the
particle-hole symmetric point the system has double degeneracy (in
fact the ground state and the low-energy excitations are all doubly
degenerate). For $\delta=B=0$, this is a direct consequence of the
system being non-ergodic, but the double degeneracy is also found for
$\delta \neq 0$ (see also below), where the system is, in fact,
ergodic. While the $\delta=B=0$ regime is in some sense pathological,
we find that the behavior of the system varies smoothly as these
parameters are varied from 0, therefore the results of the NRG
calculations for $\delta=B=0$ are also physically relevant in
spite of the system being, strictly-speaking, non-ergodic.

\subsection{Thermodynamic properties: asymmetric case}

\begin{figure}[htbp]
\centering
\includegraphics[clip,width=8cm]{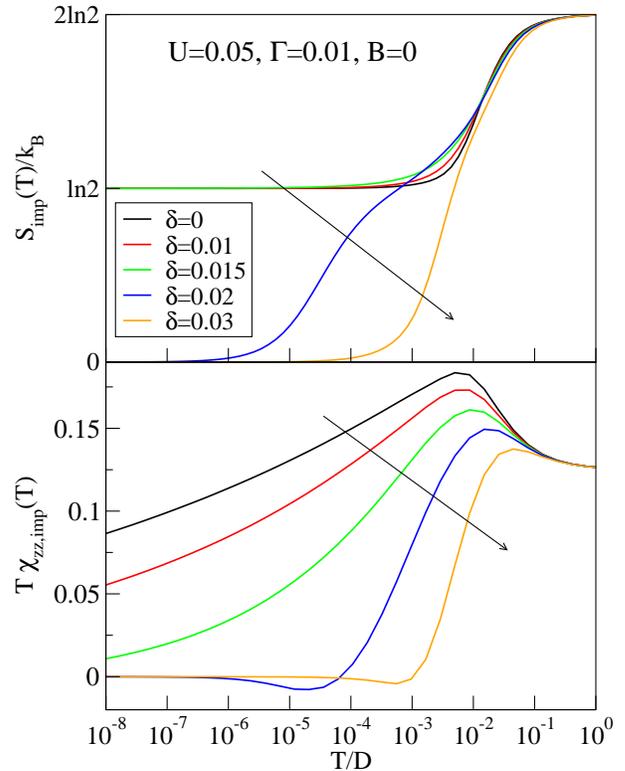}
\caption{(Color online) Thermodynamic behavior away from the
particle-hole symmetric point at $\delta=0$. The arrows indicate the
direction of increasing $\delta$. The transition point corresponds to
$|\delta| \sim U/2-c\Gamma$ with $c$ of order 1.}
\label{fig4}
\end{figure}

\begin{figure}[htbp]
\centering
\includegraphics[clip,width=7cm]{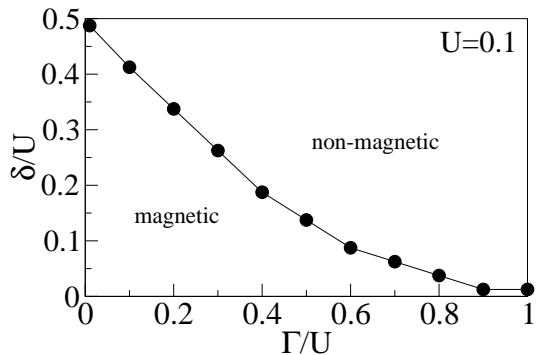}
\caption{Phase diagram in the ($\Gamma,\delta$)
plane for fixed interaction parameter $U$.
}
\label{diag}
\end{figure}

In the magnetic regime ($U/\Gamma \gtrsim 1$), the departure from the
particle-hole (p-h) symmetric point does not affect the behavior as
long as
\begin{equation}
\label{limit1}
|\delta| < U/2-c\Gamma,
\end{equation}
where $c$ is a number of order 1, at which point the fixed point is
destabilized, see Fig.~\ref{fig4}.  Notice that a similar condition
was found in the bosonization approach [see Eq.~\eqref{bozoni}] even
though the analysis was based on a large-$\Gamma$ limit. This
condition is similar to the region of existence of the ``magnetic''
local-moment fixed point in the standard SIAM
\cite{krishna1980a,krishna1980b}. In the weak-interaction regime
($U/\Gamma \lesssim 1$), the magnetic fixed point is only stable in the
immediate vicinity of the p-h symmetric point and the residual entropy
is released at low temperatures even for small $\delta$.
There thus exists a two-dimensional sheet of transition points in the
($\delta,U,\Gamma$) parameter space which separates the regimes with
or without the residual entropy. The quantum phase transition is of
the Kosterlitz-Thouless type; the cross-over between the magnetic and
non-magnetic fixed point occurs on a temperature scale which is an
exponential function of $\delta-\delta_c$, where $\delta_c$ is the
critical value of the parameter. A numerically determined phase
diagram is shown in Fig.~\ref{diag}. We emphasize again that the
existence of a phase transition is significantly different from the
behavior of the standard SIAM, where the variation of the system
properties in the ($\delta,U,\Gamma$) space is smooth and we merely
move along a line of Fermi-liquid fixed points parameterized by the
quasi-particle scattering phase shift.

By changing the on-site energy $\delta$, we tune the level occupancy
$n$. The occupancy operator $n$ is trivially related to the
$z$-component of the isospin (also known as the axial charge) operator
$i_z=1/2(n-1)$, see Eq.~\eqref{spinisospin}. In this section, we have
thus established that the magnetic fixed point is stable with respect
to the non-zero isospin field $i_z$. Due to the symmetry of the
problem, this implies that the system is also stable with respect to
moderately large isospin field $i_x$, which can be induced by the
superconducting proximity effect. We find, however, that non-zero
isospin field $i_y$ will drive the system to the non-magnetic fixed
point on the energy scale of $i_y$.

\subsection{Effects of the magnetic field}

\begin{figure}[htbp]
\centering
\includegraphics[clip,width=8cm]{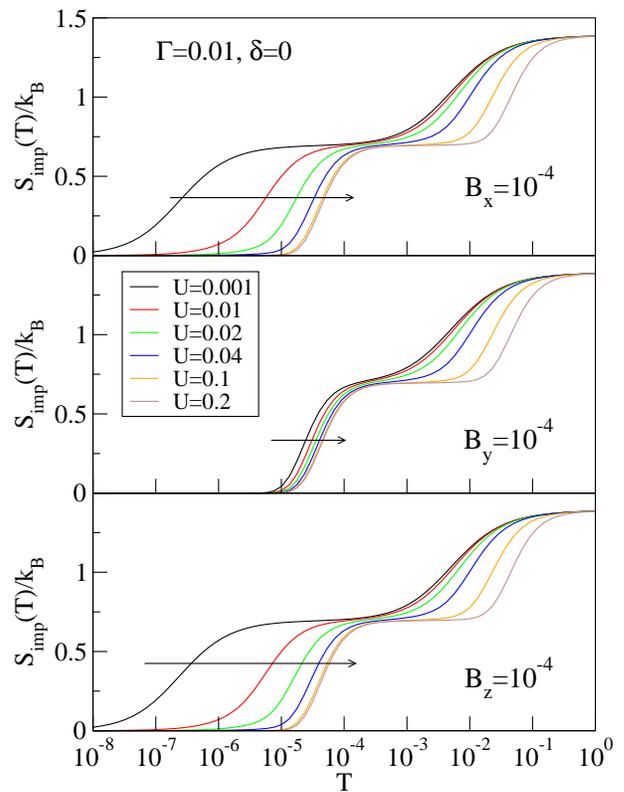}
\caption{(Color online) An impurity in an external magnetic field
along x, y, and z directions, respectively, with constant strength
$|\vc{B}|$. For each direction of the field, the parameter $U$ is
swept from the weak-interaction to the strong-interaction (Kondo)
regime. The arrows indicate the direction of increasing $U$.
}
\label{fig2u}
\end{figure}

In the presence of an external magnetic field, the entropy is released
and the effective moment goes to zero at low temperatures, see
Fig.~\ref{fig2u}. The effect of the field is direction dependent. For
a field along the ``longitudinal'' $y$ direction, the entropy is
always released on the temperature scale of $B_y$. For a field along
the ``transverse'' $x$ and $z$ directions the behavior depends on the
$U/\Gamma$ ratio, in other words, it depends on whether the impurity
is magnetic or not. For a magnetic impurity (large $U/\Gamma$ ratio)
the entropy is released on the temperature scale of $|B_\perp|$, while
for a non-magnetic impurity (small $U/\Gamma$ ratio) this only happens
on a much reduced temperature scale, see Fig.~\ref{fig2u}. This
behavior is described by the Kondo scale $T_K^-$ defined in
Eq.~\eqref{tkminus}, although the parameters $B_\perp$ and $\Gamma$
need to be rescaled by some constant factors in order to obtain full
numerical agreement.

\begin{figure}[htbp]
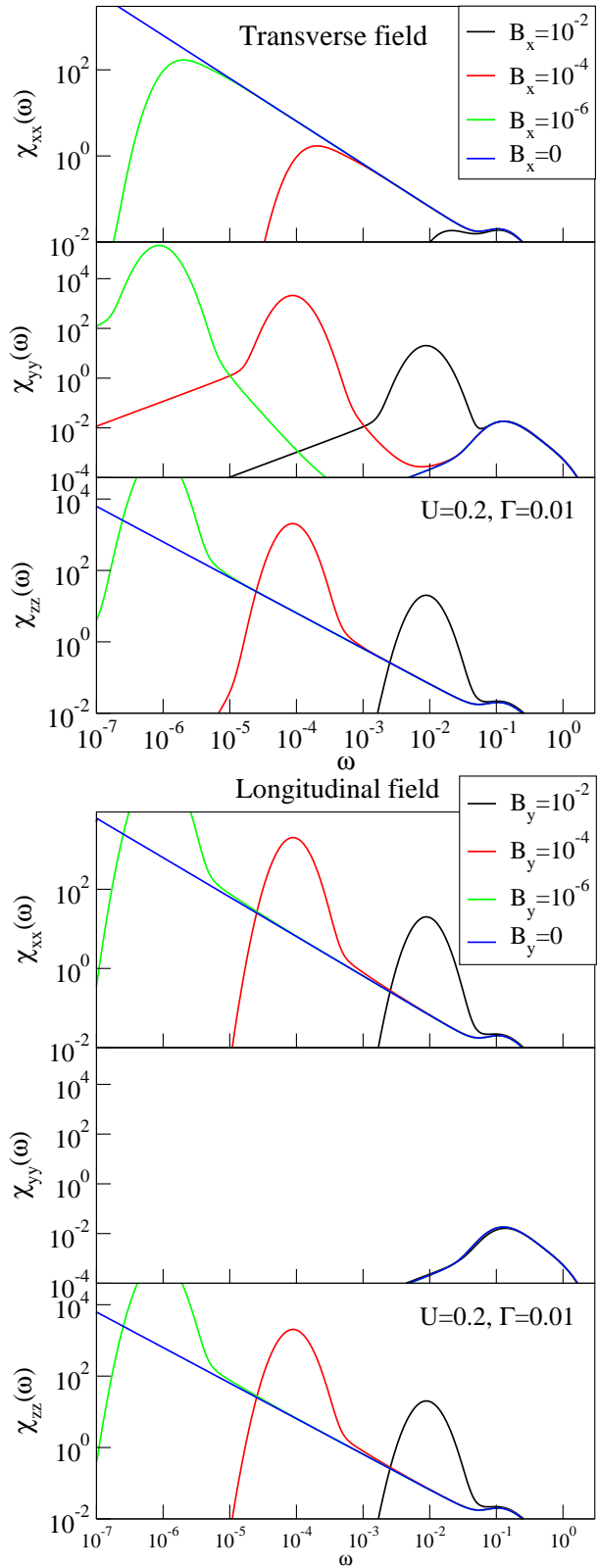

\centering
\includegraphics[clip,width=8cm]{fig8.eps}
\includegraphics[clip,width=8cm]{fig9.eps}
\caption{(Color online) Dynamical magnetic susceptibility functions
for an external magnetic field applied along the transverse ($x$)
direction (top panels) and along the longitudinal ($y$) direction
(bottom panels).}
\label{figBx}
\end{figure}

The zero-temperature dynamical magnetic susceptibility curves for zero
and non-zero magnetic field along transverse and longitudinal
directions are shown in Fig.~\ref{figBx}. In the absence of the field,
the magnetic susceptibility in the transverse directions diverges at
$\omega=0$, while the longitudinal susceptibility goes to zero. All
components of the magnetic susceptibility also feature a peak on the
frequency scale of the atomic energies (at $\sim U$, if we are in the
Kondo regime, see Figs.~\ref{figBx}). In standard SIAM, the magnetic
susceptibility peaks at $T_K$ and goes to zero at small frequencies as
a linear function of $\omega$, as mandated by the Korringa relation
for Fermi-liquid systems \cite{hewson}; the susceptibility curve also
has a peak at $\omega \sim U/2$. 

In the presence of a transverse magnetic field $B_x$, all components
of the magnetic susceptibility tensor peak on the scale of the
magnetic field and then go to zero. This is even true for the
longitudinal susceptibility which in the absence of the field goes to
zero.
For longitudinal magnetic field $B_y$, the transverse components of
the magnetic susceptibility peak on the scale of the magnetic field,
while the longitudinal susceptibility peaks on the same scale as in
the absence of the field and no further peak emerges at $\omega \sim
B_y$.
These features of the magnetic response of the system are
characteristic for impurities coupled to helical Majorana edge states
and clearly distinct from that of the standard SIAM.
We also note that the results do not seem to agree with those found
for an equivalent Kondo impurity problem in
Ref.~\onlinecite{shindou2010}. The difference can be explained in part
by the fact that the Anderson and Kondo impurity models are not fully
equivalent; see Sec.~\ref{sec4}.

\subsection{Spectral functions}

\begin{figure}[htbp]
\centering
\includegraphics[clip,width=8cm]{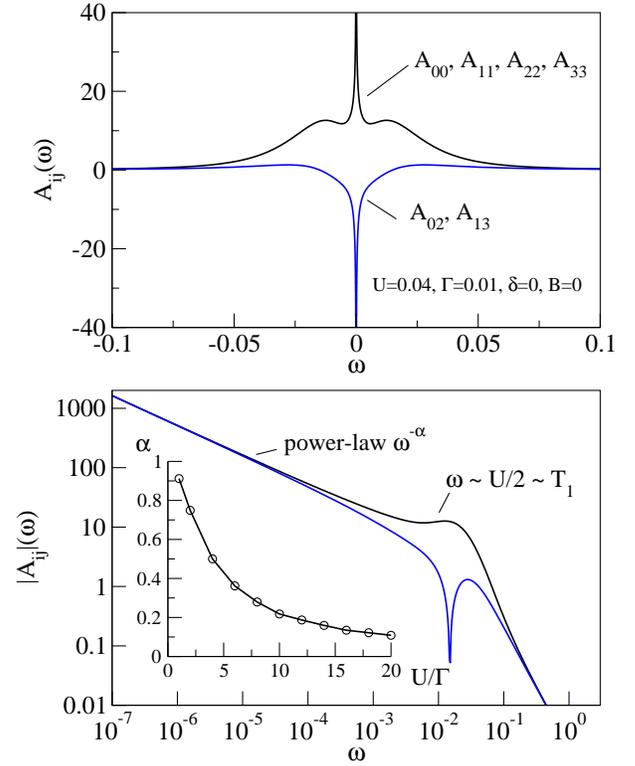}
\caption{(Color online) Spectral properties of the impurity. The
non-zero spectral functions are shown. We introduce the notation $d_0
= d^\dag_\downarrow$, $d_1 = d^\dag_\uparrow$, $d_2 =d_{\downarrow}$,
$d_3= d_{\uparrow}$ and define the spectral functions as
$A_{ij}(\omega) = -(1/\pi) \Im \langle \langle d_i ; d_j^\dag \rangle
\rangle_{\omega+i\delta}$. Non-zero anomalous spectral functions
$A_{02}$ and $A_{13}$ thus indicate the presence of pairing
correlations. The inset in the bottom panel represents the power-law
exponent as a function of the interaction strength.}
\label{fig6}
\end{figure}

The zero-temperature spectral functions of the impurity are shown in
Fig.~\ref{fig6}. In addition to the charge-fluctuation peaks at
$\omega \sim U/2$, as in the standard SIAM, one observes a sharp
resonance on the scale of $T_2$ with an inverse-power-law shape, i.e.,
the spectral function diverges. (Of course, one cannot attach a scale
to a power-law function. Instead, $T_2$ roughly corresponds to the
energy where the cross-over to the power-law behavior occurs.) This
inverse-power-law divergence replaces the Kondo resonance of the
standard SIAM, which at the lowest energy scales looks like a
parabolic peak and is finite. For strictly decoupled Majorana modes
one would expect a delta peak at zero frequency; the ``broadening''
into the inverse-power-law peak is thus an interaction effect related
to the Anderson orthogonality catastrophe physics. The exponent
depends on the interaction strength, see the inset in Fig.~\ref{fig6}.
We find that the exponent $\alpha$ is well fitted by
$\alpha=1-\frac{2}{\pi}\arctan(U/4\Gamma)$.

We also observe that the anomalous spectral functions are non-zero,
which is expected for an impurity coupled to a superconductor.

\section{Schrieffer-Wolff transformation}
\label{sec4}

We now consider a generalization of the Hamiltonian $H$ defined in
Eqs.~(\ref{eq0}-\ref{eq5}). We take the conduction-band Hamiltonian as
in the standard SIAM:
\begin{equation}
H'_3 = \sum_{k\sigma} \epsilon_k c^\dag_{k\sigma} c_{k\sigma}.
\end{equation}
This conduction band has twice the number of the degrees of freedom as
that described by Eq.~\eqref{eq5} since there are now two Majorana
modes at each $k$: $\eta_{k1\sigma}$ and $\eta_{k2\sigma}$. The
Majorana and Dirac modes are related by
\begin{equation}
\begin{split}
c_{k\sigma} = \frac{1}{\sqrt{2}} \left( \eta_{k1\sigma}
+i \eta_{k2\sigma} \right), 
&\quad
c^\dag_{k\sigma} = \frac{1}{\sqrt{2}} \left( \eta_{k1\sigma}
-i \eta_{k2\sigma} \right) \\
\eta_{k1\sigma} = \frac{1}{\sqrt{2}} \left( c_{k\sigma}
+ c_{k\sigma}^\dag \right),
&\quad
\eta_{k2\sigma} = \frac{1}{i\sqrt{2}} \left( c_{k\sigma}
- c_{k\sigma}^\dag \right).
\end{split}
\end{equation}
The hybridization term allows for unequal coupling of the Majorana
components:
\begin{equation}
\begin{split}
H'_2 &= \alpha \sum_{k\sigma} V_k (\eta_{k1\sigma} d_\sigma
+d^\dag_\sigma \eta_{k1\sigma})/\sqrt{2}\\
&+ \beta \sum_{k\sigma} V_k (-i\eta_{k2\sigma} d_\sigma
+id^\dag_\sigma \eta_{k2\sigma})/\sqrt{2},
\end{split}
\end{equation}
where $\alpha$ and $\beta$ are some real numbers. For
$\alpha=\beta=1$, this is the standard SIAM, while for $\alpha=0$,
$\beta=1$ we recover the Majorana SIAM studied in this work (up to
some differences in the notation and a factor of $\sqrt{2}$ in the
definition of $V_k$). 
We can thus smoothly interpolate between the two limiting cases. In
the following, for reasons of simplicity we assume $V_k$ to be
independent of $k$, i.e., $V_k \equiv V$. 

\newcommand{\LL}{\mathcal{L}} 
\newcommand{\OO}{\mathcal{O}}
\newcommand{\ki}{\{k_i\}}
\newcommand{\kj}{\{k_j\}}

We now proceed to perform the Schrieffer-Wolff transformation
\cite{schrieffer1966,bravyi2011} for the generalized Anderson impurity
model in the large-$U$ limit in order to derive an effective Kondo
model. The low-energy subspace consists of those many-particle states
where the impurity is singly occupied, and we denote by $P_0$ the
projection operator onto this subspace. Then $Q_0=1-P_0$ is the
projector onto the orthogonal (``high-energy'') subspace. We introduce
the superoperator 
\begin{equation}
\OO(X)=P_0 X Q_0+Q_0 X P_0.
\end{equation}
We find that $\OO(H_2')=H_2'$, thus $H_2'$ is a block-off-diagonal
operator. We next introduce the superoperator
\begin{equation}
\label{ll}
\LL(X) = \sum_{i,j} \frac{\bra{i} \OO(X) \ket{j}}{E_i - E_j}
\ket{i}\bra{j}-\text{H.c}.
\end{equation}
Here $\{ \ket{i} \}$ is an orthonormal basis of $H_0=H_1+H_3'$ such
that $H_0\ket{i}=E_i\ket{i}$. We assume that the states $\ket{i}$ in
Eq.~\eqref{ll} belong to the low-energy subspace, while $\ket{j}$
belong to the high-energy subspace (the $-\text{H.c.}$ term then
generates the remaining terms). To lowest order in hopping $aV$, one
has
\begin{equation}
S =\LL(H_2') + O(V^2),
\end{equation}
and
\begin{equation}
H_\mathrm{eff}=H_0 P_0 + \frac{1}{2} P_0 [S,H_2'] P_0 + O(V^4).
\end{equation}
We find
\begin{equation}
\begin{split}
\LL(H_2') &= \sum_{\sigma,\gamma,\ki,\kj}
\frac{\bra{\sigma\ki}H_2 \ket{\gamma\kj}}{E_{\sigma\ki}
-E_{\gamma\kj}} \\
& \ket{\sigma\ki} \bra{\gamma\kj} - \text{H.c.},
\end{split}
\end{equation}
where $\sigma \in \{\uparrow,\downarrow\}$ indexes the impurity states
in the low-energy subspace, $\gamma \in \{0,2\}$ indexes the
impurity states in the high-energy subspace, while $\ki$ and $\kj$ are
the occupancies of the conduction-band states. All terms in $H_2'$ are
such that the sets $\ki$ and $\kj$ must differ by the occupancy of a
single level. 
We obtain 
\begin{equation}
\begin{split}
S
&= \sum_{k\sigma} \frac{(\beta-\alpha)V/2}
{\epsilon_k+\epsilon}
(1-n_{-\sigma}) c^\dag_{k\sigma} d^\dag_\sigma - \text{H.c.} \\
&+ \sum_{k\sigma} \frac{(\alpha+\beta)V/2}
{\epsilon_k-\epsilon}
(1-n_{-\sigma}) c^\dag_{k\sigma} d_\sigma - \text{H.c.} \\
&+ \sum_{k\sigma} \frac{(\beta-\alpha)V/2}
{\epsilon_k+\epsilon+U}
n_{-\sigma} c^\dag_{k\sigma} d^\dag_\sigma - \text{H.c.} \\
&+ \sum_{k\sigma} \frac{(\alpha+\beta)V/2}
{\epsilon_k-\epsilon-U}
n_{-\sigma} c^\dag_{k\sigma} d_\sigma - \text{H.c.}
\end{split}
\end{equation}

The effective Hamiltonian is very complicated. A number of terms are
similar to those in the Schrieffer-Wolff transformation for the
standard SIAM, Ref.~\onlinecite{schrieffer1966}, but with coefficients
which depend on $\alpha$ and $\beta$ (and reproduce the standard
values in the $\alpha=\beta=1$ limit). Furthermore, some magnetic
anisotropy also arises. For example, the exchange-coupling terms can
be written as
\begin{equation}
H_\mathrm{ex} =\sum_{i \in \{x,y,z\}} \sum_{kk'} J_i (\Psi^\dag_{k'}S_i \Psi_{k})
(\Psi_d^\dag S_i \Psi_{d}),
\end{equation}
where $\vc{S}=\left\{S_x,S_y,S_z\right\}=(1/2)\boldsymbol{\sigma}$ are
the Pauli matrices,
\begin{equation}
\Psi_k = \begin{pmatrix}
c_{k\uparrow} \\
c_{k\downarrow}
\end{pmatrix}
\quad
\text{and}
\quad
\Psi_d = \begin{pmatrix}
d_\uparrow \\
d_\downarrow
\end{pmatrix}
\end{equation}
are field operators, and the effective Kondo exchange coupling
coefficients $J_i$ are equal to
\begin{equation}
\begin{split}
J_x=J_z =& \frac{V^2(\alpha+\beta)^2/2}{\epsilon_k-\epsilon}
+ \frac{V^2(\alpha-\beta)^2/2}{\epsilon_k+\epsilon}  \\
&- \frac{V^2(\alpha+\beta)^2/2}{\epsilon_k-\epsilon-U}
- \frac{V^2(\alpha-\beta)^2/2}{\epsilon_k+\epsilon+U}, \\
J_y =& \frac{V^2(\alpha+\beta)^2}{\epsilon_k-\epsilon}
-\frac{V^2(\alpha-\beta)^2}{\epsilon_k+\epsilon} \\
&-\frac{V^2(\alpha+\beta)^2}{\epsilon_k-\epsilon-U}
+\frac{V^2(\alpha-\beta)^2}{\epsilon_k+\epsilon+U}.
\end{split}
\end{equation}
Only in the $\alpha=\beta=1$ limit is the effective exchange
scattering isotropic. At the particle-hole symmetric point we find
\begin{equation}
\begin{split}
J_x=J_z &= \alpha\beta \left( \frac{2V^2}{\epsilon_k+U/2}
- \frac{2V^2}{\epsilon_k-U/2} \right), \\
J_y &= \frac{(\alpha^2+\beta^2)}{2} \left( \frac{2V^2}{\epsilon_k+U/2}
- \frac{2V^2}{\epsilon_k-U/2} \right).
\end{split}
\end{equation}

The direct $s$-$d$ interaction is
\begin{equation}
H_\mathrm{dir} = \sum_{kk'} \left[ W-\frac{J_z}{4}
\left(
\Psi_d^\dag \Psi_d
\right) 
\right] 
\left( \Psi_{k'}^\dag \Psi_{k} \right),
\end{equation}
where 
\begin{equation}
W = \frac{V^2 (\alpha+\beta)^2/4}{\epsilon_k-\epsilon}
+
\frac{V^2 (\alpha-\beta)^2/4}{\epsilon_k+\epsilon}.
\end{equation}

The term which can be absorbed in the Hamiltonian $H_0=H_1+H_3'$ can be
written as
\begin{equation}
H_0' = -\sum_{k\sigma} (W'+\frac{1}{2}J' n_{d,-\sigma}) n_{d,\sigma}
\end{equation}
where
\begin{equation}
\begin{split}
W' =& \frac{V^2 (\alpha+\beta)^2/4}{\epsilon_k-\epsilon}
-
\frac{V^2 (\alpha-\beta)^2/2}{\epsilon_k+\epsilon} \\
&+
\frac{V^2 (\alpha-\beta)^2/4}{\epsilon_k+\epsilon+U} \\
J' =& -\frac{V^2 (\alpha+\beta)^2/2}{\epsilon_k-\epsilon}
+ \frac{V^2 (\alpha-\beta)^2}{\epsilon_k+\epsilon} \\
&+ \frac{V^2 (\alpha+\beta)^2}{\epsilon_k-\epsilon-U}
- \frac{V^2 (\alpha-\beta)^2}{\epsilon_k+\epsilon_U}.
\end{split}
\end{equation}

The two-particle hopping term is
\begin{equation}
H_\mathrm{ch} = \frac{1}{4} \sum_{kk'\sigma} J_{2\textrm{h}}
c_{k',-\sigma}^\dag c_{k,\sigma}^\dag d_{\sigma} d_{-\sigma} +
\text{H.c.},
\end{equation}
with
\begin{equation}
J_{2\mathrm{h}} = -\frac{V^2(\alpha+\beta)^2}{\epsilon_k-\epsilon}
+\frac{V^2(\alpha+\beta)^2}{\epsilon_k-\epsilon-U}.
\end{equation}

These are, however, not all the terms which appear in the effective
Hamiltonian. Rather than enumerate all remaining contributions, we
restrict our attention only to those which maintain the single
occupancy of the impurity orbital. They can be written as:
\begin{equation}
H_\mathrm{si} = \sum_{kk'} L \left( c^\dag_{k\uparrow}
c^\dag_{k'\downarrow} - c_{k'\downarrow} c_{k\uparrow} \right) 
\left( d^\dag_\uparrow d_\downarrow -
d^\dag_\downarrow d_\uparrow \right)
\end{equation}
with
\begin{equation}
\begin{split}
L &= \frac{V^2(\beta^2-\alpha^2)/8}{\epsilon_k-\epsilon}
+ \frac{V^2(\alpha^2-\beta^2)/8}{\epsilon_k+\epsilon} \\
&+ \frac{V^2(\beta^2-\alpha^2)/8}{\epsilon_k-\epsilon-U}
+ \frac{V^2(\beta^2-\alpha^2)/8}{\epsilon_k+\epsilon+U}.
\end{split}
\end{equation}

We now discuss the $\alpha=0,\beta=1$ limit, which is relevant for
discussing an Anderson impurity coupled to the edge states of a
helical topological superconductor. The most important terms in the
effective Hamiltonian are those which affect the impurity spin degrees
of freedom:
\begin{equation}
H_\mathrm{eff} = H_\mathrm{ex} + H_\mathrm{si}.
\end{equation}
The exchange coupling constants are anisotropic:
\begin{equation}
\begin{split}
J_{x/z} &= 
\frac{V^2/2}{\epsilon_k-\epsilon}
+ \frac{V^2/2}{\epsilon_k+\epsilon}
- \frac{V^2/2}{\epsilon_k-\epsilon-U}
- \frac{V^2/2}{\epsilon_k+\epsilon+U}, \\
J_y &= 
\frac{V^2/2}{\epsilon_k-\epsilon}
- \frac{V^2/2}{\epsilon_k+\epsilon}
- \frac{V^2/2}{\epsilon_k-\epsilon-U}
+ \frac{V^2/2}{\epsilon_k+\epsilon+U}.
\end{split}
\end{equation}
At the particle-hole symmetric point this simplifies to
\begin{equation}
\begin{split}
J_{x/z} &= 0, \\
J_y &= \frac{4V^2 U}{U^2-4\epsilon_k} \approx \frac{4V^2}{U},
\end{split}
\end{equation}
where we have used $\epsilon_k \to 0$. Within the same approximation, we also have
\begin{equation}
L \approx \frac{V^2}{U}.\\
\end{equation}
We combine all terms and write
\begin{equation}
\begin{split}
H_\mathrm{eff} &= \frac{V^2}{U}
\Bigl[ 4 S_{y} s_y + \left( I^+ - I^- \right)
\left( s^+ - s^- \right) \Bigr] \\
&= \frac{4V^2}{U} (S_y-I_y) s_y.
\end{split}
\end{equation}
Here upper-case operators $I$ and $S$ correspond to the
conduction-band isospin and spin, while the lower-case operators are
those of the impurity. Furthermore, $I^+ = I_x + i I_y$, etc. We thus
conclude that in the large-$U$ limit, the impurity spin degree of
freedom couples to a mixed $S_y-I_y$ mode, which can be expressed in
terms of a single Majorana channel. This result was postulated without
derivation in Ref.~\onlinecite{shindou2010}. It is important to note,
however, that the derivation and the effective model only make sense
in the large-$U$ limit, i.e., the coupling constant $J_K = 4V^2/U$
verifies $J_K\ll V\ll U$.
The phase transition discussed in
Ref.~\onlinecite{shindou2010} occurs at a value of $\rho J_K\sim O(1)$ which is
therefore unphysical. Thus we conclude that such phase transition is 
not expected in real magnetic impurities coupled to topological superconductors.

\section{Conclusion}

We have studied the properties of the modified single-impurity
Anderson model where the impurity couples only to half of the degrees
of freedom of standard fermionic particles in the conduction-band
continuum, i.e., to Majorana fermion channels. We have shown that the
model is related to the two-level resonant model with attractive
charge-charge interaction and to the (antiferromagnetic or
ferromagnetic) anisotropic Kondo model. Two different stable
low-temperature fixed points have been identified: one corresponds to
magnetic impurities and is characterized by residual magnetic
moment and entropy, the other corresponds to non-magnetic impurities
with no residual degrees of freedom. The phase diagram separating
these two regimes has been established. We have also shown that the
magnetic field always destabilizes the magnetic fixed point, however
the response is strongly anisotropic. 

For magnetic impurities, the residual degree of freedom corresponds to
the operator $s_y+i_y$. While the charge degrees of freedom are
quenched due to electron-electron repulsion, for impurities which are
not in the extreme Kondo limit ($U \gg \Gamma$) the isospin
degree of freedom determines the quantitative aspects of the problem.
For this reason, we have shown that it is important to properly map the original Anderson
model to an effective Kondo model, rather than start by postulating a
Kondo-like model.

Finally, we would like to stress that the
predictions we have made can be put to an experimental test since
dynamical magnetic susceptibility is in principle measurable experimentally. 

\begin{acknowledgments}
R. {\v Z}. acknowledges the support of the Slovenian Research 
Agency (ARRS) under Grant No. Z1-2058 and Program P1-0044.
\end{acknowledgments}   

\appendix

\section{Complex hybridization matrix elements}
\label{appa}

For completeness, we now study the case of general complex and
spin-dependent hybridization matrix elements, that is,
the coupling Hamiltonian is written as
\begin{equation}
H_2=\sum_{k\sigma} \left( V_\sigma \eta_{k\sigma} d_\sigma
+ V_\sigma^* d^\dag_\sigma \eta_{k\sigma} \right),
\end{equation}
where $V_\sigma$ is a complex number, $V_\sigma=|V_\sigma| \exp(i
\theta_\sigma)$.

Introducing $\eta_{i\sigma}$ as in Eq.~\eqref{etaisig},
we obtain
\begin{equation}
H_2 = \sum_{k\sigma} |V_{\sigma}| \sqrt{2} i
\left( \cos\theta_\sigma \eta_{k\sigma} \eta_{2\sigma}
+ \sin\theta_\sigma \eta_{k\sigma} \eta_{1\sigma} \right).
\end{equation}
We now perform a change of basis to
\begin{equation}
\begin{split}
\xi_{1\sigma} &= \cos \theta_\sigma \eta_{1\sigma} - \sin\theta_\sigma
\eta_{2\sigma},\\
\xi_{2\sigma} &= \sin \theta_\sigma \eta_{1\sigma} + \cos\theta_\sigma
\eta_{2\sigma},
\end{split}
\end{equation}
and obtain
\begin{equation}
H_2=\sum_{k\sigma} |V_\sigma| \sqrt{2} i \eta_{k\sigma} \xi_{2\sigma}.
\end{equation}
Thus we find, again, that only two Majorana local modes
$\xi_{2\sigma}$ are coupled to the impurity, while the remaining two
modes $\xi_{1\sigma}$ are decoupled. If $\xi_{1\uparrow}
\xi_{1\downarrow}$ is rewritten in terms of the original Dirac
operators, it is found that it corresponds to a linear combination of
spin and isospin operators in the $(xy)$ plane:
\begin{equation}
\begin{split}
i \xi_{1\downarrow} \xi_{1\uparrow} = &
\sin(\theta_\downarrow-\theta_\uparrow) s_x
+\cos(\theta_\downarrow-\theta_\uparrow) s_y \\
-&\sin(\theta_\uparrow+\theta_\downarrow) i_x
+\cos(\theta_\uparrow+\theta_\downarrow) i_y.
\end{split}
\end{equation}
In the absence of spin-dependence, i.e., for
$\theta_\uparrow=\theta_\downarrow$, we obtain
\begin{equation}
i \xi_{1\downarrow} \xi_{1\uparrow} =
s_y - \sin(2\theta) i_x + \cos(2\theta) i_y,
\end{equation}
which implies that the privileged ``longitudinal'' magnetic axis is
still along the $y$ direction, however the privileged isospin axis
deviates by an angle of $2\theta$ from the $y$-axis. For
$\theta_\uparrow=\theta_\downarrow=0$ we recover the result from the
main text,
\begin{equation}
i \xi_{1\downarrow} \xi_{1\uparrow} = s_y + i_y.
\end{equation}
We remind the reader that the $z$-axis is defined by the spin
polarization of the Majorana edge modes. The $x$ and $y$-axes are thus
always defined relative to this spin quantization axis.

\end{document}